\documentclass[article]{jss_arxiv}
\usepackage[utf8]{inputenc}

\author{
Bradley C. Saul\\NoviSci, LLC \And Michael G. Hudgens\\UNC Chapel Hill
}
\title{The Calculus of M-Estimation in \proglang{R} with \pkg{geex}}

\Plainauthor{Bradley C. Saul, Michael G. Hudgens}
\Plaintitle{The Calculus of M-Estimation in R with geex}
\Shorttitle{\pkg{geex}: M-Estimation in \proglang{R}}

\Abstract{
\noindent M-estimation, or estimating equation, methods are widely
applicable for point estimation and asymptotic inference. In this paper,
we present an R package that can find roots and compute the empirical
sandwich variance estimator for any set of user-specified, unbiased
estimating equations. Examples from the M-estimation primer by
\citet{stefanski2002} demonstrate use of the software. The package also
includes a framework for finite sample, heteroscedastic, and
autocorrelation variance corrections, and a website with an extensive
collection of tutorials.
}

\Plainkeywords{empirical sandwich variance estimator, estimating equations, M-estimation, robust statistics, R}

\Address{
    Bradley C. Saul\\
  NoviSci, LLC\\
  Durham, NC\\
  E-mail: \href{mailto:bradleysaul@gmail.com}{\nolinkurl{bradleysaul@gmail.com}}\\
  URL: \url{http://www.github.com/bsaul}\\~\\
    }

\usepackage{amsmath, amsfonts, amssymb}
\usepackage{setspace}
\usepackage{graphicx}
\usepackage{multirow}
\usepackage{float}

\begin{document}

\hypertarget{introduction}{%
\section{Introduction}\label{introduction}}

M-estimation methods are general class of statistical procedures for
carrying out point estimation and asymptotic inference \citep{boos2013}.
Also known as estimating equation or estimating function methods,
M-estimation was originally developed in studying the large sample
properties of robust statistics \citep{huber2009robust}. The general
result from M-estimation theory states that if an estimator can be
expressed as the solution to an unbiased estimating equation, then under
suitable regularity conditions the estimator is asymptotically Normal
and its asymptotic variance can be consistently estimated using the
empirical sandwich estimator. Many estimators can be expressed as
solutions to unbiased estimating equations; thus M-estimation has
extensive applicability. The primer by \citet{stefanski2002}
demonstrates a variety of statistics which can be expressed as
M-estimators, including the popular method of generalized estimating
equations (GEE) for longitudinal data analysis \citep{liang1986}.

Despite the broad applicability of M-estimation, existing statistical
software packages implement M-estimators specific to particular forms of
estimating equations such as GEE. This paper introduces the package
\pkg{geex} for \proglang{R} \citep{R2017}, which can obtain point and
variance estimates from any set of unbiased estimating equations. The
analyst translates the mathematical expression of an estimating function
into an \proglang{R} function that takes unit-level data and returns a
function in terms of parameters. The package \pkg{geex} then uses
numerical routines to compute parameter estimates and the empirical
sandwich variance estimator.

This paper is outlined as follows. Section
\ref{from-m-estimation-math-to-code} reviews M-estimation theory and
outlines how \pkg{geex} translates mathematical expressions of
estimating functions into \proglang{R} syntax. Section
\ref{calculus-of-m-estimation-examples} shows several examples with
increasing complexity: three examples from \citet{stefanski2002}
(hereafter SB), GEE, and a doubly robust causal estimator
\citep{lunceford2004stratification}. All of the SB examples and several
more are available at the package website
(\url{https://bsaul.github.io/geex/}). Section
\ref{comparison-to-existing-software} compares \pkg{geex} to existing
\proglang{R} packages. Section \ref{variance-corrections} demonstrates
the variance modification feature of \pkg{geex} with examples of finite
sample corrections and autocorrelation consistent variance estimators
for correlated data. Section \ref{summary} concludes with a brief
discussion of the software's didactic utility and pragmatic
applications.

\hypertarget{from-m-estimation-math-to-code}{%
\section{From M-estimation math to
code}\label{from-m-estimation-math-to-code}}

In the basic set-up, M-estimation applies to estimators of the
\(p \times 1\) parameter
\(\theta=(\theta_1, \theta_2, \dots, \theta_p)^{\top}\) which can be
obtained as solutions to an equation of the form

\begin{equation}
\label{eq:psi}
\sum_{i = 1}^m \psi(O_i, \theta) = 0,
\end{equation}

\noindent where \(O_1, \ldots, O_m\) are \(m\) independent and
identically distributed (iid) random variables, and the function
\(\psi\) returns a vector of length \(p\) and does not depend on \(i\)
or \(m\). See SB for the case where the \(O_i\) are independent but not
necessarily identically distributed. The root of Equation \ref{eq:psi}
is referred to as an M-estimator and denoted by \(\hat \theta\).
M-estimators can be solved for analytically in some cases or computed
numerically in general. Under certain regularity conditions, the
asymptotic properties of \(\hat{\theta}\) can be derived from Taylor
series approximations, the law of large numbers, and the central limit
theorem \citep[sec.~7.2]{boos2013}. In brief, let \(\theta_0\) be the
true parameter value defined by \(\int \psi(o, \theta_0) dF(o) = 0\),
where \(F\) is the distribution function of \(O\). Let
\(\dot{\psi}(O_i, \theta) = {\partial \psi(O_i, \theta)/\partial \theta}^{\top}\),
\(A(\theta_0) = \E[-\dot{\psi}(O_1, \theta_0)]\), and
\(B(\theta_0) = \E[\psi(O_1, \theta_0)\psi(O_1, \theta_0)^{\top}]\).
Then under suitable regularity assumptions, \(\hat{\theta}\) is
consistent and asymptotically Normal, i.e.,

\begin{equation*}
\label{eq:variance}
\sqrt{m}(\hat{\theta} - \theta_0) \overset{d}{\to} N(0, V(\theta_0))  \text{ as } m \to \infty,
\end{equation*}

\noindent where
\(V(\theta_0) = A(\theta_0)^{-1} B(\theta_0) \{A(\theta_0)^{-1} \}^{\top}\).
The sandwich form of \(V(\theta_0)\) suggests several possible large
sample variance estimators. For some problems, the analytic form of
\(V(\theta_0)\) can be derived and estimators of \(\theta_0\) and other
unknowns simply plugged into \(V(\theta_0)\). Alternatively,
\(V(\theta_0)\) can be consistently estimated by the empirical sandwich
variance estimator, where the expectations in \(A(\theta)\) and
\(B(\theta)\) are replaced with their empirical counterparts. Let
\(A_i = - \dot{\psi}(O_i, \theta)|_{\theta = \hat{\theta}}\),
\(\bar{A}_m = m^{-1} \sum_{i = 1}^m A_i\),
\(B_i = \psi(O_i, \hat{\theta}) \psi(O_i, \hat{\theta})^{\top}\), and
\(\bar{B}_m = m^{-1} \sum_{i = 1}^m B_i\). The empirical sandwich
estimator of the variance of \(\hat{\theta}\) is:

\begin{equation}
\label{eq:esve}
\hat{\Sigma} = \bar{A}_m^{-1} \bar{B}_m \{\bar{A}_m^{-1}\}^{\top}/m .
\end{equation}

The \pkg{geex} package provides an application programming interface
(API) for carrying out M-estimation. The analyst provides a function,
called \code{estFUN}, corresponding to \(\psi(O_i, \theta)\) that maps
data \(O_i\) to a function of \(\theta\). Numerical derivatives
approximate \(\dot{\psi}\) so that evaluating \(\hat{\Sigma}\) is
entirely a computational exercise. No analytic derivations are required
from the analyst.

Consider estimating the population mean \(\theta = \E[Y_i]\) using the
sample mean \(\hat{\theta} = m^{-1} \sum_{i=1}^m Y_i\) of \(m\) iid
random variables \(Y_1, \dots, Y_m\). The estimator \(\hat{\theta}\) can
be expressed as the solution to the following estimating equation:

\[
\sum_{i = 1}^m (Y_i - \theta) = 0.
\]

\noindent which is equivalent to solving Equation \ref{eq:psi} where
\(O_i = Y_i\) and \(\psi(O_i, \theta) = Y_i - \theta\). An \code{estFUN}
is a translation of \(\psi\) into an \proglang{R} function whose first
argument is \code{data} and returns a function whose first argument is
\code{theta}. An \code{estFUN} corresponding to the estimating equation
for the sample mean of \(Y\) is:

\begin{verbatim}
meanFUN <- function(data){ function(theta){ data$Y - theta } } .
\end{verbatim}

The \pkg{geex} package exploits \proglang{R} as functional programming
language: functions can return and modify other functions
\citep[ch.~10]{wickham2014}. If an estimator fits into the above
framework, then the user need only specify \code{estFUN}. No other
programming is required to obtain point and variance estimates. The
remaining sections provide examples of translating \(\psi\) into an
\code{estFUN}.

\hypertarget{calculus-of-m-estimation-examples}{%
\section{Calculus of M-estimation
examples}\label{calculus-of-m-estimation-examples}}

The \pkg{geex} package can be installed from CRAN with
\code{install.packages("geex")}. The first three examples of
M-estimation from SB are presented here for demonstration. For these
examples, the data are \(O_i = \{Y_{1i}, Y_{2i}\}\) where
\(Y_1 \sim N(5, 16)\) and \(Y_2 \sim N(2, 1)\) for \(m = 100\) and are
included in the \code{geexex} dataset. Another example applies GEE,
which is elaborated on in Section \ref{variance-corrections} to
demonstrate finite sample corrections. Lastly, a doubly-robust causal
estimator of a risk difference introduces how estimating functions from
multiple models can be stacked using \pkg{geex}.

\hypertarget{example-1-sample-moments}{%
\subsection{Example 1: Sample moments}\label{example-1-sample-moments}}

The first example estimates the population mean (\(\theta_1\)) and
variance (\(\theta_2\)) of \(Y_1\). Figure \ref{ex1} shows the
estimating equations and corresponding \code{estFUN} code. The solution
to the estimating equations in Figure \ref{ex1} are the sample mean
\(\hat{\theta}_1 = m^{-1} \sum_{i = 1}^m Y_{1i}\) and sample variance
\(\hat{\theta}_2 = m^{-1} \sum_{i = 1}^m (Y_{1i} - \hat{\theta}_1)^2\).

\begin{figure}[H]
\centering
\begin{tabular}{@{\extracolsep{20pt}}lc}
 $\psi(Y_{1i}, \theta) =
 \begin{pmatrix}
  Y_{1i} - \theta_1 \\
  (Y_{1i} - \theta_1)^2 - \theta_2
 \end{pmatrix}$ &
\begin{minipage}{3in}
\begin{verbatim}
SB1_estfun <- function(data){
  Y1 <- data$Y1
  function(theta){
      c(Y1 - theta[1],
       (Y1 - theta[1])^2 - theta[2])
  }
}
\end{verbatim}
\end{minipage}
\end{tabular}
\caption{Estimating equations and \code{estFUN} for example 1.}
\label{ex1}
\end{figure}

The primary \pkg{geex} function is \code{m_estimate}, which requires two
inputs: \code{estFUN} (the \(\psi\) function), \code{data} (the data
frame containing \(O_i\) for \(i = 1, \dots, m\)). The package defaults
to \code{rootSolve::multiroot} \citep{soetaert2008practical, rootsolve}
for estimating the roots of Equation \ref{eq:psi}, though the solver
algorithm can be specified in the \code{root_control} argument. Starting
values for \code{rootSolve::multiroot} are passed via the
\code{root_control} argument; e.g.,

\begin{CodeChunk}

\begin{CodeInput}
R> library("geex")
R> results <- m_estimate(
R+     estFUN = SB1_estfun,
R+     data   = geexex,
R+     root_control = setup_root_control(start = c(1, 1)))
\end{CodeInput}
\end{CodeChunk}

The \code{m_estimate} function returns an object of the \code{S4} class
\pkg{geex}, which contains an \code{estimates} slot and \code{vcov} slot
for \(\hat{\theta}\) and \(\hat{\Sigma}\), respectively. These slots can
be accessed by the functions \code{coef} (or \code{roots}) and
\code{vcov}. The point estimates obtained for \(\theta_1\) and
\(\theta_2\) are analogous to the base \proglang{R} functions
\code{mean} and \code{var} (after multiplying by \(m-1/m\) for the
latter). SB gave a closed form for \(A(\theta_0)\) (an identity matrix)
and \(B(\theta_0)\) (not shown here) and suggest plugging in sample
moments to compute \(B(\hat{\theta})\). The maximum absolute difference
between either the point or variance estimates is 4e-11, thus
demonstrating excellent agreement between the numerical results obtained
from \pkg{geex} and the closed form solutions for this set of estimating
equations and data.

\hypertarget{example-2-ratio-estimator}{%
\subsection{Example 2: Ratio
estimator}\label{example-2-ratio-estimator}}

This example calculates a ratio estimator (Figure \ref{ex2}) and
illustrates the delta method via M-estimation. The estimating equations
target the means of \(Y_1\) and \(Y_2\), labelled \(\theta_1\) and
\(\theta_2\), as well as the estimand \(\theta_3 = \theta_1/ \theta_2\).

\begin{figure}[H]
\centering
\begin{tabular}{@{\extracolsep{20pt}}lc}
$\psi(Y_{1i}, Y_{2i}, \theta) =
\begin{pmatrix}
Y_{1i} - \theta_1 \\
Y_{2i} - \theta_2 \\
\theta_1 - \theta_3\theta_2
\end{pmatrix}$ &
\begin{minipage}{3in}
\begin{verbatim}
SB2_estfun <- function(data){
  Y1 <- data$Y1; Y2 <- data$Y2
  function(theta){
      c(Y1 - theta[1],
        Y2 - theta[2],
        theta[1] - (theta[3] * theta[2])
    )
  }
}
\end{verbatim}
\end{minipage}
\end{tabular}
\caption{Estimating equations and \code{estFUN} for example 2.}
\label{ex2}
\end{figure}

\noindent The solution to Equation \ref{eq:psi} for this \(\psi\)
function yields \(\hat{\theta}_3 = \bar{Y}_1 / \bar{Y}_2\), where
\(\bar{Y}_j\) denotes the sample mean of \(Y_{j1}, \ldots,Y_{jm}\) for
\(j=1,2\).

SB gave closed form expressions for \(A(\theta_0)\) and \(B(\theta_0)\),
into which we plug in appropriate estimates for the matrix components
and compare to the results from \pkg{geex}. The point estimates again
show excellent agreement (maximum absolute difference 4.4e-16), while
the covariance estimates differ by the third decimal (maximum absolute
difference 2e-12).

\hypertarget{example-3-delta-method-continued}{%
\subsection{Example 3: Delta method
continued}\label{example-3-delta-method-continued}}

This example extends Example 1 to again illustrate the delta method. The
estimating equations target not only the mean (\(\theta_1\)) and
variance (\(\theta_2\)) of \(Y_1\), but also the standard deviation
(\(\theta_3\)) and the log of the variance (\(\theta_4\)) of \(Y_1\).

\begin{figure}[H]
\centering
\begin{tabular}{@{\extracolsep{20pt}}lc}
$\psi(Y_{1i}, \mathbf{\theta}) =
\begin{pmatrix}
Y_{1i} - \theta_1 \\
(Y_{1i} - \theta_1)^2 - \theta_2 \\
\sqrt{\theta_2} - \theta_3 \\
\log(\theta_2) - \theta_4
\end{pmatrix}$ &
\begin{minipage}{3in}
\begin{verbatim}
SB3_estfun <- function(data){
  Y1 <- data$Y1
  function(theta){
      c(Y1 - theta[1],
       (Y1 - theta[1])^2 - theta[2],
       sqrt(theta[2]) - theta[3],
       log(theta[2]) - theta[4])
  }
}
\end{verbatim}
\end{minipage}
\end{tabular}
\label{ex3}
\caption{Estimating equations and \code{estFUN} for example 3.}
\end{figure}

SB again provided a closed form for \(A(\theta_0)\) and \(B(\theta_0)\),
which we compare to the \pkg{geex} results. The maximum absolute
difference between \pkg{geex} and the closed form estimates for both the
parameters and the covariance is 3.8e-11.

\hypertarget{example-4-generalized-estimating-equations}{%
\subsection{Example 4: Generalized estimating
equations}\label{example-4-generalized-estimating-equations}}

In their seminal paper, \citet{liang1986} introduced generalized
estimating equations (GEE) for the analysis of longitudinal or clustered
data. Let \(m\) denote the number of independent clusters. For cluster
\(i\), let \(n_i\) be the cluster size, \(Y_i\) be the \(n_i \times 1\)
outcome vector, and \(X_i\) be the \(n_i \times p\) matrix of
covariates. Let \(\mu(X_i; \theta) = \E[Y_i | X_i; \theta]\) and assume
\(\mu(X_i; \theta) = g^{-1}(X_i \theta)\), where \(g\) is some
user-specified link function. The generalized estimating equations are:

\begin{equation}
\label{gee}
\sum_{i= 1}^m \psi(O_i, \theta) = \sum_{i = 1}^m D_i^{\top} V_i^{-1} \{Y_i - \mu(X_i; \theta)\} = 0
\end{equation}

\noindent where \(O_i = \{Y_i, X_i\}\) and
\(D_i = \partial \mu(X_i; \theta)/\partial \theta\). The covariance
matrix is modeled by \(V_i = \phi W_i^{0.5} R(\alpha) W_i^{0.5}\) where
the matrix \(R(\alpha)\) is the ``working'' correlation matrix. The
example below uses an exchangeable correlation structure with
off-diagonal elements \(\alpha\). The matrix \(W_i\) is a diagonal
matrix with elements containing
\(\partial^2 \mu(X_i; \theta)/ \partial \theta^2\). Equation \ref{gee}
can be translated into an \code{estFUN} as:

\begin{CodeChunk}

\begin{CodeInput}
R> gee_estfun <- function(data, formula, family){
R+   X <- model.matrix(object = formula, data = data)
R+   Y <- model.response(model.frame(formula = formula, data = data))
R+   n <- nrow(X)
R+   function(theta, alpha, psi){
R+     mu  <- drop(family$linkinv(X %*% theta))
R+     Dt  <- crossprod(X, diag(mu, nrow = n))
R+     W   <- diag(family$variance(mu), nrow = n)
R+     R   <- matrix(alpha, nrow = n, ncol = n)
R+     diag(R) <- 1
R+     V   <- psi * (sqrt(W) %*% R %*% sqrt(W))
R+     Dt %*% solve(V, (Y - mu))
R+   }
R+ }
\end{CodeInput}
\end{CodeChunk}

This \code{estFUN} treats the correlation parameter \(\alpha\) and scale
parameter \(\phi\) as fixed, though some estimation algorithms use an
iterative procedure that alternates between estimating \(\theta_0\) and
these parameters. By customizing the root finding function, such an
algorithm can be implemented using \pkg{geex} {[}see
\code{vignette("v03_root_solvers")} for more information{]}.

We use this example to compare covariance estimates obtained from the
\code{gee} function \citep{gee}, and so do not estimate roots using
\pkg{geex}. To compute only the sandwich variance estimator, set
\code{compute_roots = FALSE} and pass estimates of \(\theta_0\) via the
\code{roots} argument. For this example, estimated roots of Equation
\ref{gee}, i.e., \(\hat{\theta}\), and estimates for \(\alpha\) and
\(\phi\) are extracted from the object returned by \code{gee}. This
example shows that an \code{estFUN} can accept additional arguments to
be passed to either the outer (\code{data}) function or the inner
(\code{theta}) function. Unlike previous examples, the independent units
are clusters (types of wool), which is specified in \code{m_estimate} by
the \code{units} argument. By default, \(m\) equals the number of rows
in the data frame.

\begin{CodeChunk}

\begin{CodeInput}
R> g <- gee::gee(breaks~tension, id=wool, data=warpbreaks,
R+               corstr="exchangeable")
\end{CodeInput}
\end{CodeChunk}

\begin{CodeChunk}

\begin{CodeInput}
R> results <- m_estimate(
R+   estFUN = gee_estfun,
R+   data  = warpbreaks,
R+   units = "wool",
R+   roots = coef(g),
R+   compute_roots = FALSE,
R+   outer_args = list(formula = breaks ~ tension,
R+                     family  = gaussian()),
R+   inner_args = list(alpha   = g$working.correlation[1,2],
R+                     psi     = g$scale))
\end{CodeInput}
\end{CodeChunk}

The maximum absolute difference between the estimated covariances
computed by \code{gee} and \pkg{geex} is 2.7e-09.

\hypertarget{example-5-doubly-robust-causal-effect-estimator}{%
\subsection{Example 5: Doubly robust causal effect
estimator}\label{example-5-doubly-robust-causal-effect-estimator}}

Estimators of causal effects often have the form:

\begin{equation}
\label{eq:causal}
\sum_{i = 1}^m \psi(O_i, \theta) = \sum_{i = 1}^m \begin{pmatrix} \psi(O_i, \nu) \\ \psi(O_i, \beta) \end{pmatrix} = 0,
\end{equation}

\noindent where \(\nu\) are parameters in nuisance model(s), such as a
propensity score model, and \(\beta\) are the target causal parameters.
Even when \(\nu\) represent parameters in common statistical models,
deriving a closed form for a sandwich variance estimator for
\(\hat{\beta}\) based on Equation \ref{eq:causal} may involve tedious
and error-prone derivative and matrix calculations \citep[e.g., see the
appendices of][ and
\citet{perez-heydrich2014assessing}]{lunceford2004stratification}. In
this example, we show how an analyst can avoid these calculations and
compute the empirical sandwich variance estimator using \pkg{geex}.

\citet{lunceford2004stratification} review several estimators of causal
effects from observational data. To demonstrate a more complicated
estimator involving multiple nuisance models, we implement the doubly
robust estimator:

\begin{equation}
\label{eq:dbr}
\hat{\Delta}_{DR} = \sum_{i = 1}^m \frac{Z_iY_i - (Z_i - \hat{e}_i) m_1(X_i, \hat{\alpha}_1)}{\hat{e}_i} - \frac{(1 - Z_i)Y_i - (Z_i - \hat{e}_i) m_0(X_i, \hat{\alpha}_0)}{1 - \hat{e}_i}.
\end{equation}

This estimator targets the average causal effect,
\(\Delta = \E[Y(1) - Y(0)]\), where \(Y(z)\) is the potential outcome
for an observational unit had it been exposed to the level \(z\) of the
binary exposure variable \(Z\). The estimated propensity score,
\(\hat{e}_i\), is the estimated probability that unit \(i\) received
\(z = 1\) and \(m_z(X_i, \hat{\alpha}_z)\) is an outcome regression
model with baseline covariates \(X_i\) and estimated paramaters
\(\hat{\alpha}_z\) for the subset of units with \(Z = z\). This
estimator has the property that if either the propensity score model or
the outcome models are correctly specified, then the solution to
Equation \ref{eq:dbr} will be a consistent and asymptotically Normal
estimator of \(\Delta\).

This estimator and its estimating equations can be translated into an
\code{estFUN} as:

\begin{CodeChunk}

\begin{CodeInput}
R> dr_estFUN <- function(data, models){
R+
R+   Z <- data$Z
R+   Y <- data$Y
R+
R+   Xe <- grab_design_matrix(
R+     data = data,
R+     rhs_formula = grab_fixed_formula(models$e))
R+   Xm0 <- grab_design_matrix(
R+     data = data,
R+     rhs_formula = grab_fixed_formula(models$m0))
R+   Xm1 <- grab_design_matrix(
R+     data = data,
R+     rhs_formula = grab_fixed_formula(models$m1))
R+
R+   e_pos  <- 1:ncol(Xe)
R+   m0_pos <- (max(e_pos) + 1):(max(e_pos) + ncol(Xm0))
R+   m1_pos <- (max(m0_pos) + 1):(max(m0_pos) + ncol(Xm1))
R+
R+   e_scores  <- grab_psiFUN(models$e, data)
R+   m0_scores <- grab_psiFUN(models$m0, data)
R+   m1_scores <- grab_psiFUN(models$m1, data)
R+
R+   function(theta){
R+     e  <- plogis(Xe %*% theta[e_pos])
R+     m0 <- Xm0 %*% theta[m0_pos]
R+     m1 <- Xm1 %*% theta[m1_pos]
R+     rd_hat <- (Z*Y - (Z - e) * m1) / e -
R+       ((1 - Z) * Y - (Z - e) * m0) / (1 - e)
R+     c(e_scores(theta[e_pos]),
R+       m0_scores(theta[m0_pos]) * (Z == 0),
R+       m1_scores(theta[m1_pos]) * (Z == 1),
R+       rd_hat - theta[length(theta)])
R+   }
R+ }
\end{CodeInput}
\end{CodeChunk}

\noindent This \code{estFUN} presumes that the user will pass a list
containing fitted model objects for the three nuisance models: the
propensity score model and one regression model for each treatment
group. The functions \code{grab_design_matrix} and
\code{grab_fixed_formula} are \pkg{geex} utilities for extracting
relevant pieces of a model object. The function \code{grab_psiFUN}
converts a fitted model object to an estimating function; for example,
for a \code{glm} object, \code{grab_psiFUN} uses the \code{data} to
create a \code{function} of \code{theta} corresponding to the
generalized linear model score function. The \code{m_estimate} function
can be wrapped in another function, wherein nuisance models are fit and
passed to \code{m_estimate}.

\begin{CodeChunk}

\begin{CodeInput}
R> estimate_dr <- function(data, propensity_formula, outcome_formula){
R+   e_model  <- glm(propensity_formula, data = data, family = binomial)
R+   m0_model <- glm(outcome_formula, subset = (Z == 0), data = data)
R+   m1_model <- glm(outcome_formula, subset = (Z == 1), data = data)
R+   models <- list(e = e_model, m0 = m0_model, m1 = m1_model)
R+   nparms <- sum(unlist(lapply(models, function(x) length(coef(x))))) + 1
R+
R+   m_estimate(
R+     estFUN = dr_estFUN,
R+     data   = data,
R+     root_control = setup_root_control(start = rep(0, nparms)),
R+     outer_args = list(models = models))
R+ }
\end{CodeInput}
\end{CodeChunk}

The following code provides a function to replicate the simulation
settings of \citet{lunceford2004stratification}.

\begin{CodeChunk}

\begin{CodeInput}
R> library("mvtnorm")
R> tau_0 <- c(-1, -1, 1, 1)
R> tau_1 <- tau_0 * -1
R> Sigma_X3 <- matrix(
R+    c(1, 0.5, -0.5, -0.5,
R+      0.5, 1, -0.5, -0.5,
R+      -0.5, -0.5, 1, 0.5,
R+      -0.5, -0.5, 0.5, 1), ncol = 4, byrow = TRUE)
R>
R> gen_data <- function(n, beta, nu, xi){
R+   X3 <- rbinom(n, 1, prob = 0.2)
R+   V3 <- rbinom(n, 1, prob = (0.75 * X3 + (0.25 * (1 - X3))))
R+   hold <- rmvnorm(n,  mean = rep(0, 4), Sigma_X3)
R+   colnames(hold) <- c("X1", "V1", "X2", "V2")
R+   hold <- cbind(hold, X3, V3)
R+   hold <- apply(hold, 1, function(x){
R+     x[1:4] <- x[1:4] + tau_1^(x[5]) * tau_0^(1 - x[5])
R+     x
R+   })
R+   hold <- t(hold)[, c("X1", "X2", "X3", "V1", "V2", "V3")]
R+   X <- cbind(Int = 1, hold)
R+   Z <- rbinom(n, 1, prob = plogis(X[, 1:4] %*% beta))
R+   X <- cbind(X[, 1:4], Z, X[, 5:7])
R+   data.frame(
R+     Y = X %*% c(nu, xi) + rnorm(n),
R+     X[ , -1])
R+ }
\end{CodeInput}
\end{CodeChunk}

To show that \code{estimate_dr} correctly computes
\(\hat{\Delta}_{DR}\), the results from \pkg{geex} can be compared to
computing \(\hat{\Delta}_{DR}\) ``by hand'' for a simulated dataset.

\begin{CodeChunk}

\begin{CodeInput}
R> dt <- gen_data(n    = 1000,
R+                beta = c(0, 0.6, -0.6, 0.6),
R+                nu   =  c(0, -1, 1, -1, 2),
R+                xi   = c(-1, 1, 1))
R> geex_results <- estimate_dr(dt, Z ~ X1 + X2 + X3, Y ~ X1 + X2 + X3)
\end{CodeInput}
\end{CodeChunk}

\begin{CodeChunk}

\begin{CodeInput}
R> e  <- predict(glm(Z ~ X1 + X2 + X3, data = dt, family = "binomial"),
R+               type = "response")
R> m0 <- predict(glm(Y ~ X1 + X2 + X3, data = dt, subset = Z==0),
R+               newdata = dt)
R> m1 <- predict(glm(Y ~ X1 + X2 + X3, data = dt, subset = Z==1),
R+               newdata = dt)
R> del_hat <- with(dt, mean( (Z * Y - (Z - e) * m1) / e)) -
R+   with(dt, mean(((1 - Z) * Y - (Z - e) * m0) / (1 - e)))
\end{CodeInput}
\end{CodeChunk}

The maximum absolute difference between \code{coef(geex_results)[13]}
and \code{del_hat} is 1.4e-09.

\hypertarget{comparison-to-existing-software}{%
\section{Comparison to existing
software}\label{comparison-to-existing-software}}

The above examples demonstrate the basic utility of the \pkg{geex}
package and the power of R's functional programming capability. The
\pkg{gmm} package \citep{chausse2010computing} computes generalized
methods of moments and generalized empirical likelihoods, estimation
strategies similar to M-estimation, using user-defined functions like
\pkg{geex}. To our knowledge, \pkg{geex} is the first \proglang{R}
package to create an extensible API for any estimator that is the
solution to estimating equations in the form of Equation \ref{eq:psi}.
Existing \proglang{R} packages such as \pkg{gee} \citep{gee},
\pkg{geepack} \citep{geepack}, and \pkg{geeM} \citep{geeM} solve for
parameters in a GEE framework. Other packages such as \pkg{fastM}
\citep{fastm} and \pkg{smoothmest} \citep{smoothmest} implement
M-estimators for specific use cases.

For computing a sandwich variance estimator, \pkg{geex} is similar to
the popular \code{sandwich} package
\citep{zeileis2004econometric, zeileis2006object}, which computes the
empirical sandwich variance estimator from modelling methods such as
\code{lm}, \code{glm}, \code{gam}, \code{survreg}, and others. For
comparison to the exposition herein, the infrastructure of
\pkg{sandwich} is explained in \citet{zeileis2006object}. Advantages of
\pkg{geex} compared to \pkg{sandwich} include: (i) for custom
applications, a user only needs to specify a single \code{estFUN}
function as opposed to both the \code{bread} and \code{estfun}
functions; (ii) as demonstrated in the examples above, the syntax of an
\code{estFUN} may closely resemble the mathematical expression of the
corresponding estimating function; (iii) estimating functions from
multiple models are easily stacked; and (iv) point estimates can be
obtained. The precision and computational speed of point and variance
estimation in \pkg{geex}, however, depends on numerical approximations
rather than analytic expressions.

To compare \pkg{sandwich} and \pkg{geex}, consider estimating
\(\hat{\Sigma}\) for the \(\theta\) parameters in the following simple
linear model contained in the \code{geexex} data:
\(Y_4 = \theta_1 + \theta_2 X_1 + \theta_3 X_2 + \epsilon\), where
\(\epsilon \sim N(0, 1)\). The estimating equation for \(\theta\) in
this model can be expressed in an \code{estFUN} as:

\begin{CodeChunk}

\begin{CodeInput}
R> lm_estfun <- function(data){
R+   X <- cbind(1, data[["X1"]], data[["X2"]])
R+   Y <- data[["Y4"]]
R+   function(theta){
R+     crossprod(X, Y - X %*% theta)
R+   }
R+ }
\end{CodeInput}
\end{CodeChunk}

\noindent Then \(\hat{\theta}\) and \(\hat{\Sigma}\) can be computed in
\pkg{geex}:

\begin{CodeChunk}

\begin{CodeInput}
R> results <- m_estimate(
R+   estFUN = lm_estfun,
R+   data  = geexex,
R+   root_control = setup_root_control(start = c(0, 0, 0)))
\end{CodeInput}
\end{CodeChunk}

\noindent or from the \code{lm} and \code{sandwich} functions:

\begin{CodeChunk}

\begin{CodeInput}
R> fm <- lm(Y4 ~ X1 + X2, data = geexex)
R> sand_vcov <- sandwich::sandwich(fm)
\end{CodeInput}
\end{CodeChunk}

The results are virtually identical (maximum absolute difference
1.4e-12). The \code{lm}/\code{sandwich} option is faster
computationally, but \pkg{geex} can be sped up by, for example, changing
the options of the derivative function via \code{deriv_control} or
computing \(\hat{\Sigma}\) using the parameter estimates from \code{lm}.
While \pkg{geex} will never replace computationally optimized modelling
functions such as \code{lm}, the important difference is that \pkg{geex}
lays bare the estimating function used, which is both a powerful
didactic tool as well as a programming advantage when developing custom
estimating functions.

\hypertarget{variance-corrections}{%
\section{Variance corrections}\label{variance-corrections}}

The standard empirical sandwich variance estimator is known to perform
poorly in certain situations. In small samples, \(\hat{\Sigma}\) will
tend to underestimate the variance of \(\hat{\theta}\) \citep{fay2001}.
When observational units are not independent and/or do not share the
same variance, consistent variance estimators can be obtained by
modifying how \(B(\theta_0)\) is estimated. The next two examples
demonstrate using \pkg{geex} for finite sample and autocorrelation
corrections, respectively.

\hypertarget{finite-sample-correction}{%
\subsection{Finite sample correction}\label{finite-sample-correction}}

Particularly in the context of GEE, many authors \citep[e.g.,
see][]{paul2014, li2014} have proposed corrections that modify
components of \(\hat{\Sigma}\) and/or by assuming \(\hat{\theta}\)
follows a \(t\) (or \(F\)), as opposed to Normal, distribution with some
estimated degrees of freedom. Many of the proposed corrections somehow
modify a combination of the \(A_i\), \(\bar{A}_m\), \(B_i\), or
\(\bar{B}_m\) matrices.

Users may specify functions that utilize these matrices to form
corrections within \pkg{geex}. A finite sample correction function
requires at least the argument \code{components}, which is an \code{S4}
object with slots for the \code{A} (\(= \sum_i A_i\)) matrix, \code{A_i}
(a list of all \(m\) \(A_i\) matrices), the \code{B} (\(= \sum_i B_i\))
matrix, \code{B_i} (a list of all \(m\) \(B_i\) matrices), and
\code{ee_i} (a list of the observed estimating function values for all
\(m\) units). Additional arguments may also be specified, as shown in
the example. The \pkg{geex} package includes the bias correction and
degrees of freedom corrections proposed by \citet{fay2001} in the
\code{fay_bias_correction} and \code{fay_df_correction} functions
respectively. The following demonstrates the construction and use of the
bias correction. \citet{fay2001} proposed the modified variance
estimator
\(\hat{\Sigma}^{bc}(b) = \bar{A}_m^{-1} \bar{B}_m^{bc}(b) \{\bar{A}_m^{-1}\}^{\top}/m\),
where:

\begin{equation*}
\label{eq:bc}
B^{bc}_m(b) = \sum_{i = 1}^m H_i(b) B_i H_i(b)^{\top},
\end{equation*}

\begin{equation*}
\label{eq:H}
H_i(b) = \{1 - \min(b, \{A_i \bar{A}_m^{-1}\}_{jj}) \}^{-1/2},
\end{equation*}

\noindent and \(W_{jj}\) denotes the \(jj\) element of a matrix \(W\).
When \(\{A_i \bar{A}_m^{-1}\}_{jj}\) is close to 1, the adjustment to
\(\hat{\Sigma}^{bc}(b)\) may be extreme, and the constant \(b\) is
chosen by the analyst to limit over adjustments. The bias corrected
estimator \(\hat{\Sigma}^{bc}(b)\) can be implemented in \pkg{geex} by
the following function:

\begin{CodeChunk}

\begin{CodeInput}
R> bias_correction <- function(components, b){
R+   A   <- grab_bread(components)
R+   A_i <- grab_bread_list(components)
R+   B_i <- grab_meat_list(components)
R+   Ainv <- solve(A)
R+
R+   H_i <- lapply(A_i, function(m){
R+     diag( (1 - pmin(b, diag(m %*% Ainv) ) )^(-0.5) )
R+   })
R+
R+   Bbc_i <- lapply(seq_along(B_i), function(i){
R+     H_i[[i]] %*% B_i[[i]] %*% H_i[[i]]
R+   })
R+
R+   Bbc <- compute_sum_of_list(Bbc_i)
R+   compute_sigma(A = A, B = Bbc)
R+ }
\end{CodeInput}
\end{CodeChunk}

The \code{compute_sum_of_list} sums over a list of matrices, while the
\code{compute_sigma(A, B)} function simply computes
\(A^{-1} B \{A^{-1}\}^{\top}\). To use this bias correction, the
\code{m_estimate} function accepts a named list of corrections to
perform. Each element of the list is a \code{correct_control} \code{S4}
object that can be created with the helper function \code{correction},
which accepts the argument \code{FUN} (the correction function) plus any
arguments passed to \code{FUN} besides \code{components}; e.g.,

\begin{CodeChunk}

\begin{CodeInput}
R> results <- m_estimate(
R+   estFUN = gee_estfun, data  = warpbreaks,
R+   units = "wool", roots = coef(g), compute_roots = FALSE,
R+   outer_args = list(formula = breaks ~ tension,
R+                     family  = gaussian(link = "identity")),
R+   inner_args = list(alpha   = g$working.correlation[1,2],
R+                     psi     = g$scale),
R+   corrections = list(
R+    bias_correction_.1 = correction(FUN = bias_correction, b = .1),
R+    bias_correction_.3 = correction(FUN = bias_correction, b = .3)))
\end{CodeInput}
\end{CodeChunk}

In the \pkg{geex} output, the slot \code{corrections} contains a list of
the results of computing each item in the \code{corrections}, which can
be accessed with the \code{get_corrections} function. The corrections of
\citet{fay2001} are also implemented in the \pkg{saws} package
\citep{fay2001}. Comparing the \pkg{geex} results to the results of the
\code{saws::geeUOmega} function, the maximum absolute difference for any
of the corrected estimated covariance matrices is 3.8e-09.

\hypertarget{newey-west-autocorrelation-correction}{%
\subsection{Newey-West autocorrelation
correction}\label{newey-west-autocorrelation-correction}}

When error terms are dependent, as in time series data, \(\E[B]\) is
challenging to estimate \citep{zeileis2004econometric}. A solution is to
estimate \(B\) using the pairwise sum,

\[
\hat{B}_{AC} = \sum_{i,j = 1}^m  w_{|i - j|} \psi(O_i; \hat{\theta})\psi(O_j; \hat{\theta})^{\intercal},
\]

where \(w_{|i - j|}\) is a vector of weights that often reflect
decreasing autocorrelation as the distance between \(i\) and \(j\)
increases. Many authors have proposed ways of computing weights
\citep[see for
example,][]{white1984nonlinear, newey1987a-simple, andrews1991heteroskedasticity, lumley1999weighted}.

To illustrate autocorrelation correction using \pkg{geex}, we implement
the Newey-West correction (without pre-whitening) and compare to the
\code{NeweyWest} function in \pkg{sandwich}
\citep{zeileis2004econometric}. The example is taken from the
\code{NeweyWest} documentation.

\begin{CodeChunk}

\begin{CodeInput}
R> x <- sin(1:100)
R> y <- 1 + x + rnorm(100)
R> dt <- data.frame(x = x, y = y)
R> fm <- lm(y ~ x)
R>
R> lm_estfun <- function(data){
R+   X <- cbind(1, data[["x"]])
R+   Y <- data[["y"]]
R+   function(theta){
R+     crossprod(X, Y - X %*% theta)
R+   }
R+ }
R>
R> nwFUN <- function(i, j, lag){
R+    ifelse(abs(i -j) <= lag, 1 - abs(i - j) / (lag + 1), 0)
R+ }
R>
R> nw_correction <- function(components, lag){
R+   A   <- grab_bread(components)
R+   ee  <- grab_ee_list(components)
R+   Bac <- compute_pairwise_sum_of_list(ee, .wFUN = nwFUN, lag = lag)
R+   compute_sigma(A = A, B = Bac)
R+ }
R>
R> results <- m_estimate(
R+   estFUN = lm_estfun,
R+   data   = dt,
R+   root_control = setup_root_control(start = c(0, 0)),
R+   corrections = list(
R+     NW_correction = correction(FUN = nw_correction, lag = 1)))
R>
R> get_corrections(results)[[1]]
\end{CodeInput}

\begin{CodeOutput}
            [,1]        [,2]
[1,] 0.010555254 0.003304559
[2,] 0.003304559 0.023758823
\end{CodeOutput}

\begin{CodeInput}
R> sandwich::NeweyWest(fm, lag = 1, prewhite = FALSE)
\end{CodeInput}

\begin{CodeOutput}
            (Intercept)           x
(Intercept) 0.010555254 0.003304559
x           0.003304559 0.023758823
\end{CodeOutput}
\end{CodeChunk}

The function \code{lm_estfun} is essentially the same as the previous
comparison to \pkg{sandwich} in Section
\ref{comparison-to-existing-software}. The function \code{nw_correction}
performs the Newey-West adjustment using \code{nwFUN} which computes the
Newey-West weights for lag \(L\),

\[
w_{|i - j|} = 1 - \frac{|i - j|}{L + 1}.
\]

The function \code{grab_ee_list} returns the list of observed estimating
functions, \(\psi(O_i, \hat{\theta})\), from the
\code{sandwich_components} object. The utility function
\code{compute_pairwise_sum_of_list} computes \(\hat{B}_{AC}\) using
either (but not both) a fixed vector (argument \code{.w}) of weights or
a function of \code{i} and \code{j} (argument \code{.wFUN}), which may
include additional arguments such as \code{lag}, as in this case. For
this example, \pkg{geex} and \pkg{sandwich} return nearly identical
results.

\hypertarget{summary}{%
\section{Summary}\label{summary}}

This paper demonstrates how M-estimators and finite sample corrections
can be transparently implemented in \pkg{geex}. The package website
(\url{https://bsaul.github.io/geex/}) showcases many examples of
M-estimation including instrumental variables, sample quantiles, robust
regression, generalized linear models, and more. A valuable feature of
M-estimators is that estimating functions corresponding to parameters
from multiple models may be combined, or ``stacked,'' in a single set of
estimating functions. The \pkg{geex} package makes it easy to stack
estimating functions for the target parameters with estimating functions
from each of the component models, as shown in the package vignette
\code{v06_causal_example}. Indeed, the software was motivated by causal
inference problems \citep{saul2017upstream} where target causal
parameters are functions of parameters in multiple models.

The theory of M-estimation is broadly applicable, yet existing
\proglang{R} packages only implement particular classes of M-estimators.
With its functional programming capabilities, \proglang{R} routines can
be more general. The \pkg{geex} framework epitomizes the extensible
nature of M-estimators and explicitly translates the estimating function
\(\psi\) into a corresponding \code{estFUN}. In this way, \pkg{geex}
should be useful for practitioners developing M-estimators, as well as
students learning estimating equation theory.

\section*{Acknowledgments}

The Causal Inference with Interference research group at UNC provided
helpful feedback throughout this project. Brian Barkley, in particular,
contributed and tested the software throughout its development. This
work was partially supported by NIH grant R01 AI085073. The content is
solely the responsibility of the authors and does not necessarily
represent the official views of the National Institutes of Health.

\renewcommand\refname{References}
\bibliography{geex}

\end{document}